\documentclass{josis}
\usepackage{hyperref}
\usepackage{booktabs}
\usepackage{stmaryrd}
\usepackage[T1]{fontenc}
\usepackage[utf8]{inputenc}
\usepackage{cite}

\usepackage{algorithm}
\usepackage{algpseudocode}

\usepackage[table]{xcolor}
\usepackage{amssymb,amsmath}

\makeatletter
\let\UrlSpecialsOld\UrlSpecials
\def\UrlSpecials{\UrlSpecialsOld\do\/{\Url@slash}\do\_{\Url@underscore}}%
\def\Url@slash{\@ifnextchar/{\kern-.11em\mathchar47\kern-.2em}%
    {\kern-.0em\mathchar47\kern-.08em\penalty\UrlBigBreakPenalty}}
\def\Url@underscore{\nfss@text{\leavevmode \kern.06em\vbox{\hrule\@width.3em}}}
\makeatother

\hypersetup{
colorlinks=true,
linkcolor=black,
citecolor=black,
urlcolor=black
} 

\runningauthor{\begin{minipage}{.9\textwidth}\centering Ballari, Siabato, Claramunt\end{minipage}}
\runningtitle{Development of OGDIs in Latin America}

\begin{document}

\title{On the development of open geographical data infrastructures in Latin America: progress and challenges
}

\author{Daniela Ballari}\affil{Instituto de Estudios de Régimen Seccional del Ecuador (IERSE), Facultad de Ciencia y Tecnología, Universidad del Azuay,  Ecuador}
\author{Willington Siabato}\affil{Universidad Nacional de Colombia, Department of Geography, Colombia}

\author{Christophe Claramunt}\affil{Naval Academy Research Institute, Lanvéoc-Poulmic, France}

\author{Felix Mata}\author{Roberto Zagal}\affil{Mobile Computing Laboratory, National Polytechnic Institute, Mexico}

\author{Rodolfo Franco}\affil{Universidad Distrital Francisco José de Caldas, Colombia}

\maketitle

\begingroup
\raggedright 
\setlength{\parindent}{0pt} 
\sloppy
\endgroup

\vspace{0.5cm}
\noindent\textcolor{red}{Uncorrected proof accepted for publication in the Journal of Spatial Information Science (JOSIS). Licensed under Creative Commons Attribution 3.0 License CC.}
\vspace{0.5cm}

\begin{abstract}
Open data initiatives and infrastructures play an essential role in favoring better data access, participation, and transparency in government operations and decision-making. Open Geographical Data Infrastructures (OGDIs) allow citizens to access and scrutinize government and public data, thereby enhancing accountability and evidence-based decision-making. This encourages citizen engagement and participation in public affairs and offers researchers, non-governmental organizations, civil society, and business sectors novel opportunities to analyze and disseminate large amounts of geographical data and to address social, urban, and environmental challenges. In Latin America, while recent open government agendas have shown an inclination towards transparency, citizen participation, and collaboration, only a limited number of OGDIs allow unrestricted use and re-use of their data. Given the region's cultural, social, and economic disparities, there is a contrasting digital divide that significantly impacts how OGDIs are being developed. Therefore, this paper analyses recent progress in developing OGDIs in Latin America, technological gaps, and open geographical data initiatives. The main results denote an early development of OGDIs in the region. Nevertheless, this opens the door for the timely involvement of citizens and non-government sectors to share needs, experiences, knowledge, and expertise, as well as to address a transboundary research agenda. Challenges are discussed from multiple perspectives: data, methodological, governmental and readiness, and potential impact. This analysis is aimed at researchers, policymakers, and practitioners interested in the specific challenges and progress of OGDIs in Latin America, while also contributing to the global conversation on best practices and lessons learned in implementing OGDIs across different contexts. 
\end{abstract}

\section{Introduction}
Over the past years, the extensive worldwide development of Geographical data Infrastructures (GDI) has drastically changed the way geographical data is managed and shared, particularly as digital tools and platforms have made data more accessible to a wider array of users \cite{giuliani_bringing_2017}. Volunteered Geographical Information (VGI) and crowdsourcing applications, generally developed over the Web and mobile applications, have played a key role in shaping the emerging concept of Open Geographical Data Infrastructures (OGDIs), defined as infrastructures that facilitate the management and sharing of geographical data in an open manner, encompassing both governmental repositories and more decentralized citizen-driven initiatives. This includes formal GDIs and ad-hoc project-based platforms that allow for the collection and dissemination of geographical data by a wide range of users, including the public \cite{sui_citizen_2013,goodchild_citizens_2007}. The differences between OGDIs and conventional GDIs hinge on their components, access and integration protocols, and the roles of multi-stakeholders involved in the data lifecycle.

The concept of OGDI significantly expands the notion of GDI (also known as Spatial Data Infrastructures - SDI). We suggest that the openness of GDIs can be understood as an evolutionary stage succeeding the three well-known stages defined so far \cite{mulder_status_2020}: a producer-based GDI with a focus on data provision; followed by a process-based GDI with a focus on standard web-based services for discovery, access, and visualization of data; and as third stage a user-based GDI with a focus on the users’ needs. Openness, in this context, arises from the open paradigm and principles \cite{open_knowledge_foundation_defining_2015} and it is not limited to freely accessible data. In fact, it involves a predominant role of multi-stakeholders (i.e citizens, policymakers, and businesses), in addition to the conventional role of government, for re-using data, generating new data, and actively participating in data governance. The main purpose of OGDIs is to enhance open geographical data access to favor its availability without restrictions on use, re-use, or redistribution. A valuable objective of OGDIs is to promote the implementation of best practices and management procedures that involve citizens in decision-making processes. OGDIs clearly foster novel pathways for decision-making processes that are close to the citizens and accountable.

While conventional GDIs may emphasize controlled data-sharing with predefined users, OGDIs broaden this perspective by facilitating public access to data, often integrating VGI and citizen science contributions \cite{nunes_citizen_2013,heipke_crowdsourcing_2010}. Such infrastructures often involve both persistent governmental data repositories and ephemeral citizen-driven platforms that enable data sharing through open initiatives and project-based collaborations \cite{sui_citizen_2013,zook_volunteered_2010}. Making geographical data accessible to all users typically involves ensuring the availability, accessibility, and interoperability of geographical data \cite{vancauwenberghe_assessing_2018,izdebski_open_2021-1}. Therefore, it emphasizes the use of common standards and formats to ensure that different data sources and systems can communicate and work together among the general public, researchers, policymakers, and businesses. The diversity of actors and domains of application requires the development of appropriate methodologies and design approaches for a successful implementation that fully involves citizens and produces reliable OGDIs \cite{claramunt_geomatics_2023}.

However, the transition from conventional GDI to Open GDIs is not without challenges. While a wide range of open, volunteer-based, and collaborative infrastructures are emerging, questions arise regarding the quality, governance, and long-term sustainability of these, particularly when involving citizen science projects or crowd-sourced data. OGDIs emphasize the use of common standards and formats to ensure interoperability, but the diversity of data sources and actors involved in these initiatives also brings forth issues related to data accuracy, updating, and usability across different contexts. Additionally, unequal access to technological resources in certain regions, especially in developing countries, may perpetuate digital divides rather than eliminate them.

In Latin America, while recent open government initiatives and agendas have shown an inclination towards transparency, citizen participation, and collaboration \cite{espinosa_what_2022,martinez_estudio_2022-1,vancauwenberghe_assessing_2018}, key challenges still remain. These include ensuring data quality, fostering inclusivity, sustaining volunteer engagement, aligning with technical standards, and addressing legal and privacy concerns. Most conventional GDI primarily focus on geographical data visualization. Unfortunately, only a limited number of them allow data downloading for subsequent analysis, and even fewer operate under open licenses. There is also a lack of standard and common practices that hampers cross-fertilization and successful exchanges at the methodological level. Given the region's cultural, social, and economic disparities, there is a contrasting digital divide \cite{cpidea_diagnostico_2013,hyman_survey_2003}, which significantly impacts how OGDIs are being developed across the region. There is a need to identify the models and the local policies that will shape the methodological and technological frameworks on which OGDIs should be based. While the underlying technologies may remain consistent with global standards, their application must be tailored to address the unique challenges and requirements of the region.

The focus on Latin America offers a unique perspective and opportunity for developing OGDI in this region. Analyzing the development of OGDIs within this context can provide insights into how these challenges can be addressed using a multi user-centric and data-driven approach, although not limited to purely data exchange issues but also the generation of added value and knowledge. 

Therefore, the objective of this paper is to showcase the Latin American region by analyzing the recent progress in developing OGDIs, technological gaps among countries and open geographical data initiatives. The OGDI progress in the region is stated from scientific literature and citizen open initiatives reports, from which challenges are discussed as potential for fostering innovation, citizen engagement, digital transformation, and data-driven initiatives under the open geographical data umbrella. This content is directed towards researchers, policymakers, and practitioners interested in the specific challenges and advancements of OGDIs in Latin America, while also contributing to the global discourse on best practices and lessons learned in the implementation of OGDIs across diverse contexts.

The rest of the paper is structured as follows. Section 2 briefly introduces OGDI principles and recent development trends. Section 3 develops a specific focus on the progress of Latin American OGDIs. Finally, section 4 outlines future challenges, and Section 5 summarizes the findings.

\section{OGDI principles and development}
Nowadays, the emerging and increasingly pervasive digital age has also extended access to and engagement with geographical data to new and diverse user communities. Therefore, OGDIs build on this evolution but represent a fundamental shift in how data is shared, moving from a government-centric model to one that is open to multiple stakeholders, including citizens, non-governmental organizations, policymakers and business actors. This openness creates opportunities for democratizing data but also raises challenges in terms of managing diverse data sources and ensuring consistent quality across different user communities that are playing an increasingly participatory role \cite{claramunt_geomatics_2023}. 

Behind the openness of data to non-expert users, there is the assumption that “knowledge is open if anyone is free to access, use, modify, and share it — subject, at most, to measures that preserve provenance and openness” \cite{open_knowledge_foundation_defining_2015}. The few principles ruling the open data paradigm \cite{open_data_handbook_what_2015} can be also useful for geographical data: 

\renewcommand{\labelitemi}{--}

\begin{itemize}
    \item Data availability and access: “the data must be available as a whole and at no more than a reasonable reproduction cost, preferably by downloading over the internet. The data must also be available in a convenient and modifiable form”;
    \item     Re-use and redistribution: “the data must be provided under terms that permit re-use and redistribution including the intermixing with other datasets;
    \item Universal participation: “everyone must be able to use, re-use and redistribute - there should be no discrimination against fields of endeavor or persons or groups. For example, ‘non-commercial’ restrictions that would prevent ‘commercial’ use, or restrictions of use for certain purposes (e.g. only in education), are not allowed”.
\end{itemize}

This is not the only attempt to define open data. For instance, another approach that expands on these ideas is the Open Data Charter Principles \cite{open_data_charter_odc_2023}: 

\begin{itemize}
    \item Open by default;
    \item Timely and comprehensive;
    \item Accessible and usable;
    \item Comparable and interoperable;
    \item For improved governance and citizen engagement;
    \item For inclusive development and innovation.
\end{itemize}

These principles reflect a shift toward democratizing data access, where the focus is not just on the data itself, but also on its re-use, repurposing, and adaptation by a wider range of actors beyond traditional users. From a geographical standpoint, the application of these principles within OGDIs presents both opportunities and challenges. While the ideals of open access and transparency are often promoted, in practice, the governance of these infrastructures can be fraught with tensions between different stakeholder groups, each with their own expectations and priorities.

The multi-stakeholder model reflects the transition from government-controlled infrastructures to platforms designed to foster widespread collaboration and innovation across sectors. This means that OGDIs are conceptually open to all key stakeholders for open data provision, as well as the governance and implementation of the infrastructure. The stakeholders include, besides the government, citizens, research institutions, and businesses and non-governmental actors \cite{vancauwenberghe_assessing_2018}. Thus, it is not only about open data access (free without restrictions) but also “organizing and governing the infrastructure openly, enabling and stimulating the participation of non-government actors” \cite{izdebski_open_2021-1}. A distinction should be made between end users, citizens, thematic experts, and decision-makers as OGDIs can contribute to these different user dimensions and from these different perspectives. However, the engagement of these diverse stakeholders also introduces complexities in terms of data literacy, interoperability, and governance. Data literacy plays a significant role in allowing end users, especially non-geographical data experts, to unlock the full potential of open data and be able to access, create, understand, analyze, and use geographical data \cite{montes_issues_2019}. Specific domains might have different ways of implementing OGDIs at different scales and levels of precision. The benefits of OGDIs range from transparency and government efficiency to public participation, economic growth, and innovation \cite{singleton_establishing_2016,vancauwenberghe_assessing_2018}. Last but not least, OGDIs can foster regional and global collaboration by facilitating data sharing \cite{giuliani_bringing_2017}, transboundary collaboration \cite{de_keyser_integrating_2023} and supporting Sustainable Development Goals (e.g. \cite{ballari_satellite_2023,ilie_monitoring_2019,rizvi_data_2020}).

Although implementing interoperable infrastructures and standard services for data access, GDIs did not necessarily originate as open platforms (i.e. following open data principles); many were designed as data-sharing mechanisms among government and public institutions \cite{sieber_open_2019}. OGDIs are expected to be intentionally developed for openness purposes, following the open principles, to stimulate innovation, encourage the development of new applications and services, and foster economic growth. OGDIs can be deliberately established through administrative or governance initiatives to promote transparency, improve the quality of service delivery, facilitate citizen engagement in decision-making, and leverage academic research and educational initiatives. Hence, data assets that previously were not in the public domain, or under restrictive licenses, are now released for free to use and re-use in either public or non-public applications \cite{singleton_establishing_2016}. Opening geographical data does not simply mean applying an open license to existing datasets, but also involves adopting policies, standards, and allocating human resources specifically focused on geographical data use, sharing, and distribution following the open principles \cite{sieber_open_2019}.

However, the implementation of OGDIs presents diverse challenges and experiences worldwide. For instance, several countries, such as the United States and Canada, have made significant strides in developing OGDIs that foster citizen engagement and collaboration. In the United States, initiatives like Data.gov have successfully opened vast datasets to the public, enabling various stakeholders to access and utilize geographical information for innovative solutions \cite{us_government_datagov_nodate}. Conversely, experiences in some developing countries reveal obstacles, such as inadequate infrastructure, limited data quality, and political resistance that hinder the establishment of effective OGDIs. For example, in parts of Africa, many GDIs remain largely governmental, focusing on internal use rather than promoting public participation or open data sharing \cite{guigoz_spatial_2017}. These contrasting cases illustrate the varied landscape of OGDIs worldwide and highlight the necessity for tailored approaches that consider regional contexts, governance structures, and the specific needs of diverse user groups. Overall, while OGDIs have the potential to foster global collaboration, they must be critically assessed to understand the successes and challenges faced by different regions in implementing these infrastructures \cite{begany_understanding_2021,canares_enhancing_2016}.

\section{Recent progress in Latin America}
This section provides an overview of the recent progress of OGDI in Latin America from three standpoints. First, we present one of the major challenges for OGDI development in the region: the digital divide. Second, we assess the status of OGDI based on scientific literature and reports from civil society open initiatives, with a specific focus on the Latin American region and the context of geographical data. Third, we broaden the scope to examine the Latin American landscape through the lens of global open data initiatives. Although these initiatives do not specifically focus on geographical data, they are relevant for establishing the general open data framework in the region. These three standpoints allow us to establish an analytical framework to analyze the recent progress of OGDI in Latin America.

\subsection{Digital divide}
The Latin American region is a vast and diverse group of countries that cover a large part of the American continent. Although cultural roots differ significantly, there is a growing need to adopt a more balanced approach to economic, social, and technological development \cite{eclac_2023}. These characteristics can exacerbate the risk of digital divides across the region \cite{pazmino-sarango_assessing_2022}, which in turn can limit and condition the development and effectiveness of OGDIs. Specifically, disparities in access to technology and data literacy among various populations can restrict participation in the creation and utilization of OGDIs, ultimately affecting their inclusivity and impact. 

The digital “modernity islands” on implementing open data and OGDIs have already been reported. It is well known that access to open geographical data remains uneven worldwide \cite{sieber_open_2019}, and Latin America is no exception. The technological gaps among countries in the context of the public sector digital transformation have been reported by the GovTech initiative of the Work Bank \cite{world_bank_govtech_2022}, which focused on a whole-of-government approach to public sector modernization. From the 2022 report, we selected and mapped three indexes related to the open data principles to illustrate the digital divide landscape in the region (see Figure 1). They are (i) the maturity of online public service portals from a citizen and universal accessibility design perspective (Public Service Delivery Index – PSDI); (ii) the availability of public participation platforms, citizen feedback mechanisms, open data, and open government portals (Digital Citizen Engagement Index - DCEI); and (iii) the availability of institutional strategies, laws and regulations, digital skills, and innovation policies and programs, to foster digital public sector modernization (GovTech Enablers Index - GTEI). The indexes ranked government tech practices from A (very high) to D (low), with A for “GovTech leaders”; B for “Significant focus on GovTech”, C for “Some focus on GovTech”, and D for “Minimal focus on GovTech”. For Latin America, the PSDI showed a relative homogeneity in the maturity of digital services among countries. However, the other two indexes showed critical gaps concerning citizen engagement and institutional enablers.

\begin{figure}[tbh]
\centering
\includegraphics[width=\textwidth]{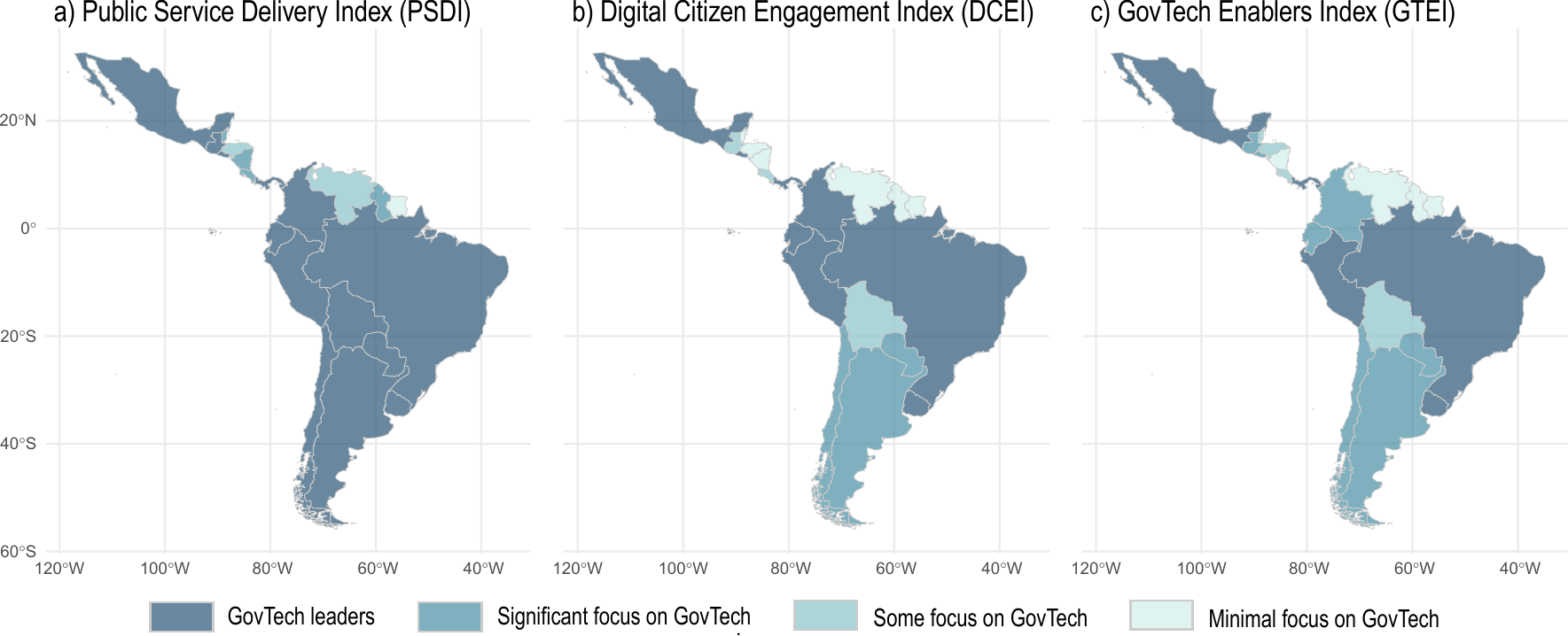}
\caption{{Selected indexes related to the open data principles for the Latin American region: a) Public Service Delivery Index – PSDI; b) Digital Citizen Engagement Index – DCEI; and c) GovTech Enablers Index – GTEI.}}\label{fig:1}
\end{figure}

\subsection{OGDI experiences in Latin America}
Few studies in the geographical data field have highlighted the regional gap for GDIs and OGDIs. Vilches \& Ballari \cite{vilches_unveiling_2020}, based on a web-based survey about GDIs in the region, revealed various levels of maturity in adopting information technology trends within GDIs (i.e. big data, semantic web, mobile devices, internet of things, or volunteer geographical information). Although some trends in GDI implementations were reported, most were still under development, and only a few were already available to public users. Among thirteen countries, the ones that reported more implemented information technology trends were Argentina, Chile, Colombia, and Ecuador. This disparity underscores not only the varying capacity of different nations to adopt technological innovations but also the potential implications for regional competitiveness in the development of OGDIs. Although the study did not focus on open infrastructures, we refer to it since it shows efforts from the Latin American community to integrate technological trends within the conventional GDIs. This, in turn, may indicates a sign of willingness to adopt open geographical data policies and fosters innovation in the future. However, the mere presence of technological trends does not guarantee effective implementation or impact, necessitating a more nuanced analysis of the socio-political and economic contexts that facilitate or hinder these advancements.

Bruzza et al. \cite{BRUZZA_2020} also showed regional gaps in open geographical data for e-government services. Argentina, Colombia, and Uruguay, followed by Mexico and Cuba, ranked higher in this regard. Based on a systematic review, a recent and notable increase in the use of geographical data for e-government was identified. On the other hand, Ecuador and Peru showed signs of underdevelopment in open government. These findings highlight a critical divide in governance capabilities and citizen engagement across the region, suggesting that efforts to enhance transparency and service delivery are unevenly distributed. The differences in the country rankings between the two mentioned studies relate to the used methodologies, thus emphasizing the need for standardized metrics to evaluate the maturity of GDIs and OGDIs.

From an open data availability perspective, there are incipient initiatives opening geographical data and addressing the challenges to integrate them. Despite extensive lists of available geoportals in Latin America offering geographical data visualization and access (e.g. \cite{franco_geoportales_2016}, most GDIs primarily provide data visualization, and some of them allow downloading data for further analysis, with restrictions on use and re-use \cite{sieber_open_2019}. This is an important limitation for achieving the vision of open data. Additionally, only a few are actually registered as open data portals. For instance, the Open Data Inception Initiative \cite{open_data_inception} provides a detailed worldwide overview of open data portals. As of December 2024, 3,744 records have been added to this dataset. Of the 397 general records for the Latin American and Caribbean region, only 56 belong to the geographical thematic category (out of 609 global). Most of them provide access to the GDIs of National Mapping Agencies and national and local public sector (47), and a minor representation is from universities, businesses, associations, and other organizations (9). While access to open data is improving in the region, the amount of open geographical datasets is still very modest. This raises concerns about the actual impact of these datasets on decision-making processes and public participation, as a lack of diversity in data sources may lead to bias or incomplete analyses.

It is worth mentioning the previous works conducted in Mexico which identified certain issues about open data sources for public health, air quality, and crime \cite{Mendez_spatio-temporal_2022,zagal_flores_geo-social_2022,hernandez_analysis_2021,mata_mobile_2016}. First, some open data did not include geographical data; however, such data might be highly feasible to be georeferenced. For instance, in the health field, georeferenced data is relevant to allow a precise understanding of the spatial distribution of diseases \cite{secretariat_of_healt_2024}. Certain small-population localities were not accounted for, and much of this information was restricted by legal aspects, hindering the ability to focus on vulnerable population groups in terms of health, public safety, and the environment. These barriers point to systemic issues within data governance that limit access to critical information, particularly for marginalized communities that could benefit the most from geographic insights. Furthermore, discrepancies exist between identifiers for regional and local administrative delineations across different open data sources. For instance, this issue was found in data concerning water consumption \cite{sacmex} or the registry of various businesses defining economic activities \cite{national_statistical}. Moreover, McKeen et al. \cite{mckeen_high-resolution_2023}, while creating open data for the distribution of residential populations across forty countries in Latin America and the Caribbean, found a lack of alignment between transboundary governmental statistics. This misalignment complicates cross-border analysis and collaboration, which are essential for addressing regional challenges that transcend national boundaries. 

While numerous open geographical datasets may exist at the local, regional, and national levels, they are often not easily discoverable due to the absence of integrated meta-search engines with OGDIs. Therefore, despite the growing availability of open geographical datasets, challenges persist in standardizing information from various territorial delineations, sources and themes, and in making it readily discoverable for end-users. This highlights a critical need for innovative solutions that not only enhance data accessibility but also ensure the relevance and usability of the data provided.

\subsection{Latin American participation in global open data initiatives}
\sloppy
In global initiatives and studies, it can be observed that Latin America shows significant engagement with open data initiatives; however, this varies among sectors \cite{fumega_open_2019}, and only a few of them address geographical data. Some initiatives showing the increasing importance of open data in the region described below. 
\sloppy
Since its inception in 2015, of the 97 worldwide national and subnational administrations that have signed and adopted the principles of the International Open Data Charter, 11 national governments and 35 local governments are from Latin America: Argentina, Chile, Colombia, Costa Rica, El Salvador, Guatemala, Honduras, Mexico, Panama, Paraguay, and Uruguay; a total of 46 representing the 47.2\% of the global signatories \cite{open_data_charter_odc_2023}.
\sloppy
More recently, the Latin American Open Data Initiative (ILDA) \cite{ilda_latin} launched the 2020 Latin American and Caribbean edition of the Open Data Barometer \cite{davies_global_2022,fumega_barometro_2020}. Although neither applied to OGDIs nor exclusively focused on geographical data, this report clearly assessed the state of data openness. The report for Latin America concluded that, although the region has made improvements in open data, it still requires significant or radical transformation. Such a transformation is essential not only for enhancing data accessibility but also for addressing systemic inequalities that hinder the equitable distribution of open data benefits. However, the region risks stagnation, compromising the potential contribution to more robust democracies, services, and sustainable development. Regarding readiness, there is an explicit government commitment to releasing open data and developing national open data portals. The highest scores are found in implementation, though at very different speeds. In the impact dimension, the region has not yet reaped the expected benefits \cite{fumega_barometro_2020}. This gap between intention and outcome suggests a need for more robust evaluation mechanisms to assess the real-world implications of open data initiatives. The top five countries with the most consolidated open data policies and technical leadership are Uruguay, Argentina, Colombia, Brazil, and Mexico \cite{fumega_barometro_2020}. The 2020 report also shows that Latin American countries have progressively improved the evaluation results in the six barometers conducted since 2013.
\sloppy
Although not recently updated, another international initiative is the Open Data Impact Map \cite{open_data_for_development}. It is a public database of organizations that showcase open data sources from 90 countries worldwide; 16 are Latin American and Caribbean countries, with Mexico, Argentina, Brazil, Uruguay, and Colombia the ones with the most registered organizations, confirming the trend of the ILDA barometer. Among different sectors such as governance, finance, communications, education, or health, IT and geographic sectors were identified as the ones that most use open data. According to this report, open geographical data is mainly used to develop products and services, such as web and mobile applications and data analytics, as well as organizational optimization, such as strategic and competitive information. This demonstrates the growing recognition of open data as a valuable asset for economic development, yet it raises concerns about the inclusivity of these benefits for a wide range of communities. Other less-reported uses include advocacy and research, which underscore the importance of collaboration among stakeholders, government, business sector, and citizens \cite{fumega_open_2019}. This collaboration is critical to ensuring that open data initiatives are not only implemented but also actively contribute to societal improvement and informed decision-making. 

\section{Challenges to address in Latin America}
As an analytical framework to summarize the ongoing challenges in the Latin American context, we have extended the OGDI framework for assessing readiness, data, and impact provided by Vancauwenberghe et al. \cite{vancauwenberghe_assessing_2018} to also include the methodological and governmental dimensions that play a critical role in the successful design, implementation and delivery of open geographical data. Therefore, the analyzed dimensions are: 
\sloppy
\begin{itemize}
    \item The \textbf{data dimension} involves the availability, accessibility, and generation of geographical data by different users inside and outside public administration, as well as the services implemented for metadata, discovery, visualization, and download. The data should be openly available to anyone, free of charge, and openly licensed \cite{vancauwenberghe_assessing_2018}.
   
    \item The \textbf{methodological dimension} denotes novel modeling approaches and technologies to acquire, integrate, and update open geographical datasets coming from different sources \cite{BRUZZA_2020} and multi-stakeholders. It favors the integrating of large structured and unstructured data at various scales and levels of abstraction, while standards and metadata with a focus on openness are not always available or implemented \cite{martinez_analysis_2023}.
     
    \item The \textbf{impact dimension} focuses on the benefits of using geographical data within and beyond the government to make informed decisions and create new products and services. The benefits should be continuously evaluated in all dimensions (economic, environmental, social, and technological, among others), and the beneficiaries and actors should objectively learn from them.    
    
    \item The \textbf{governmental and readiness dimension} involves developing and implementing OGDIs under the umbrella of legal frameworks, guidelines, policies, and political willingness for openness \cite{vancauwenberghe_assessing_2018,barenque_implementacion_2021,aguerre_open_2024,restrepo_scoping_2023}. Although readiness involves both government and non-government actors, we highlight the governmental role, and willingness for fostering policy formulation and the posterior application of such policies.
\end{itemize}

Some of these dimensions share similar backgrounds, experiences and challenges with the conventional GDI, but the openness and multi-stakeholders integrations create new challenges. Furthermore, the involved OGDI multi-stakeholders (government, citizen, business) are discussed as a complementary dimension along with the other ones. Following, we provide an overview and assessment of recent progress, limitations, and the remaining challenges that need to be addressed. Figure 2 provides a summary of the main challenges introduced in this section.

\begin{figure}[tbh]
\centering
\includegraphics[width=\textwidth]{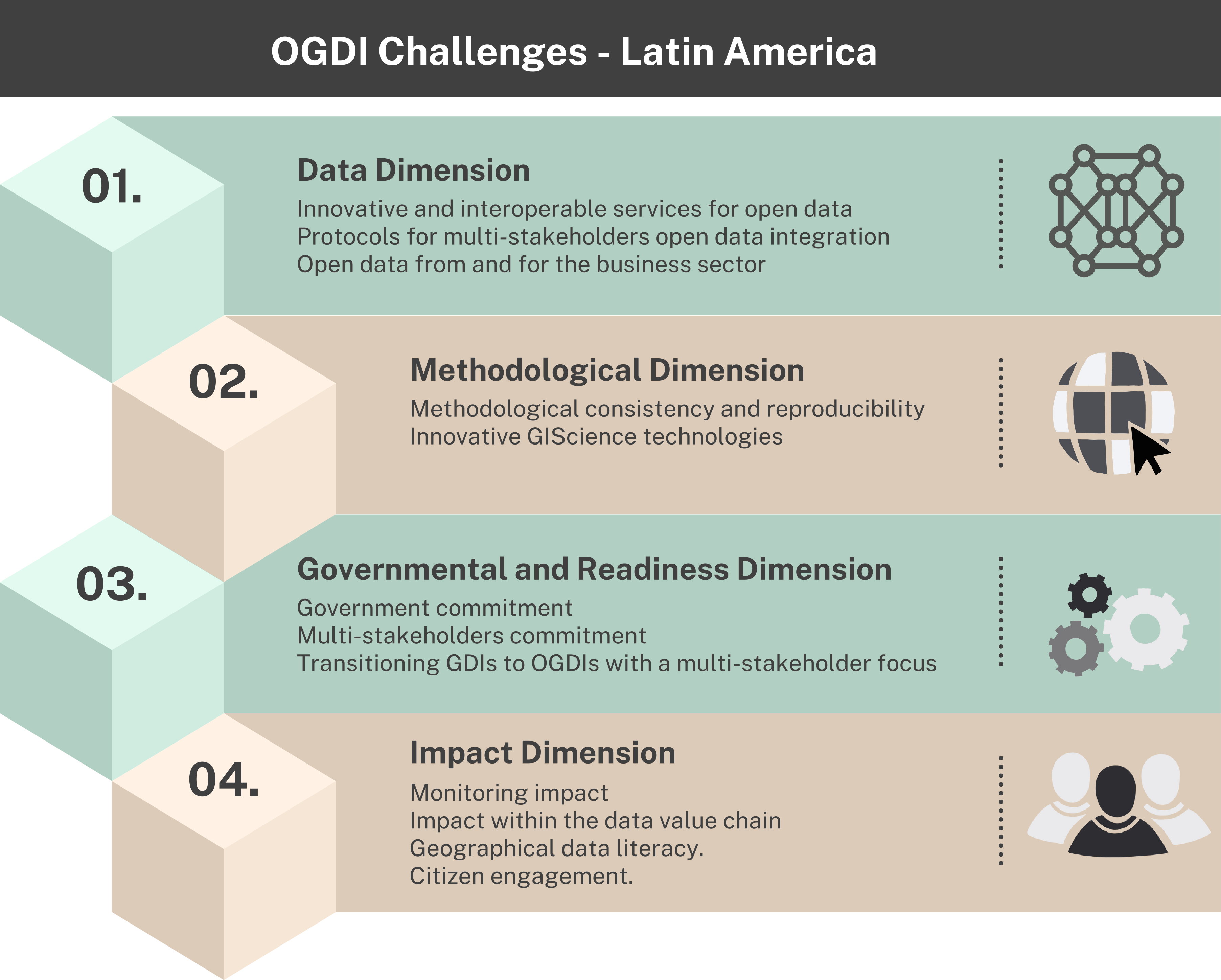}
\caption{Main challenges to address OGDIs in Latin America.}\label{fig:2}
\end{figure}

\subsection{Data dimension}
\textbf{Innovative and interoperable services for open data}. Although conventional GDIs are already well established in Latin America, the data dimension remains the most developed dimension among the discussed ones. However, efforts still need to be made to fully consider the potential of accounting for open data. There is a critical need for innovative solutions that not only enhance services for data accessibility but also ensure the relevance and usability of the data adapted to multi-stakeholders. Standardized and interoperable services are already in place, as shown in Figure 1 - Index SPDI \cite{world_bank_govtech_2022}. However, most services are from the government sector and developed by geographical data experts, with incipient (or even non-existent) citizen and non-government integration and participation. The use of OGDI data for e-government services has significantly risen across Latin America in recent years, framed within the Open Government and Open Data programs, as well as data exchange and interoperability platforms. However, users find it more challenging to locate the geographical data they need, and formal and organizational structures have thus far proven ineffective in facilitating this process \cite{BRUZZA_2020}. They also did not completely address issues for interoperating open data that are sometimes available in external repositories and under different standards \cite{martinez_analysis_2023}. Meta-search engines can play a role in the integration of these external repositories, providing a unified platform for data access, sharing and discovery. However, the effectiveness of these meta-search engines in bridging gaps in geographical data access is contingent upon their ability to aggregate high-quality datasets and ensure they meet standards of reliability and relevance. Maintaining the quality and innovation of OGDI data and services in the long term raises major technical infrastructure, data management issues, and organizational challenges \cite{martinez_estudio_2022-1}. The challenge in this regard is, in some cases, the lack of familiarity among government personnel with OGDI technological trends \cite{vilches_unveiling_2020}. Additionally, adaptation to the fast-moving technological and telecommunication trends that support OGDI development raises concerns about the need for ongoing training and adaptation of services. Therefore, more efforts need to be made in order to make open data easily findable and usable in an interoperable manner. Since building usable services and freely open data with quality requirements are not without cost to the government \cite{singleton_establishing_2016}, this is a significant challenge to address. For these, Latin American governments should recognize that such a cost is a long-term investment that can generate wealth for every nation.

\textbf{Protocols for multi-stakeholders open data integration}. The availability of open public datasets in Latin America, especially from citizen data, is still developing \cite{restrepo_scoping_2023}. The already existing open datasets are generally under-utilized because of methodological differences and lack of alignment with government statistics \cite{mckeen_high-resolution_2023}. For such an alignment, it is essential to avoid unequal representation of populations and regions by paying particular attention to gender and inclusive dimensions, ensuring the participation of all individuals in society \cite{restrepo_scoping_2023}. Thus, the integration of reliable observational data requires formulating appropriate protocols that consider the diverse actors involved in data sampling and validation \cite{jacquin_citizen_2023,wilson_open_2019}. The development of protocols should also improve the quality and curation of data \cite{restrepo_scoping_2023}. For instance, Martinez et al. \cite{martinez_analysis_2023} proposed an implementation guide for the quality of open health data after evaluating open data portals from six countries in Latin America, highlighting the need for collaboration among data providers. Geographical experts could use and extend these previous experiences and efforts for the context of geographical data. Last but not least, OGDI should address privacy and security issues to safeguard and protect individuals' data.

\textbf{Open data from and for the business sector}. The integration of the business sector with new strategies could incentivize investment and profitability \cite{fumega_open_2019}. Urgent action must be maintained with this sector to understand its needs and link them with the generation and use of open data. Many business sectors still need to learn about the benefits of integrating open data and geographical information technologies within their business.

\subsection{Methodological dimension}
\textbf{Methodological consistency and reproducibility.} One major challenge is to address the diversity in conceptual, geographical, temporal, and semantic dimensions that complicates methodological consistency and interoperability \cite{mata_geographic_2007}. The methods applied at the design and implementation levels are most often not well-defined, leading to reengineering problems for maintenance and reusability. Thus, OGDIs should be associated with the scientific expertise and reproducibility principles that support implementation of consistent methods and efficient use and re-use. 

\textbf{Innovative technologies.} There are undoubtedly many innovative applications in which GIScience and citizen science can be combined to create practical solutions for the context of OGDI. The expected increase in the availability of interactive, real-time, and affordable OGDIs will offer many opportunities for citizen scientists to collaborate with academic scientists, the business sector, and the government. Therefore, complementary and innovative technologies, such as well-designed Human-Computer interfaces (HCI), need to be developed to enhance and simplify user integration and interactions, implement rating and feedback systems, and boost learning processes and capacity building. Geographical artificial intelligence also has the potential to enhance citizen and entrepreneur participation by suggesting and prioritizing activities based on their interests and location and considering previous experience and engagement \cite{claramunt_geomatics_2023}. In this regard, it is essential to analyze the expected impact of collaborative approaches and Generative Artificial Intelligence on developing OGDIs. Appropriate gamification and rewards as motivational and engaging factors, which have been proven effective in several fields, are still to be developed in the context of geographical applications, as well as elements to guarantee transparency throughout all processes. 

\textbf{Additional research questions.} Additionally, to the previous challenges, further questions also arise: What types of open geographical data are being utilized within OGDIs in Latin America, and what methods and models are commonly employed to analyze and integrate this data? Are there emerging open geographical data models and infrastructures that have been successfully implemented across different countries in Latin America, and how can these models be adapted for broader use? What is the current level of awareness and expertise regarding the maintenance of OGDIs in Latin America, and what strategies are in place to ensure the sustainability and continuous updating of the open data provided? What are the underlying scientific processes, general processes, and implementations associated with citizen science activities in the context of OGDIs so that they can be replicated and their outcomes better understood? To what degree can a concept of methodological observational framework be developed to observe current OGDI developments and practices? 

\subsection{Governmental and readiness dimension}
\textbf{Government commitment.} Explicit Latin America´s government commitment to the open data agenda \cite{martinez_analysis_2023} and to establish national portals is evident \cite{fumega_barometro_2020}, as shown by the good performance of the Public Service Delivery Index – PSDI \cite{world_bank_govtech_2022} (Figure 1). Nevertheless, there is still a notable lack of emphasis on open data concerning geographical data, which delays the achievement of this dimension. Furthermore, this dimension plays a significant role in Latin America because a formal commitment is challenging to create and maintain \cite{aguerre_open_2024}, whether political instability is frequent. Such a commitment can be strengthened by embracing an open data model as a fundamental aspect of modern governance by prioritizing the creation, access, and distribution of both public and citizen data. Therefore, resources, data integration models, policies, and an organizational structure are needed to institutionalize the integration of citizen data with public data \cite{jacquin_citizen_2023}. By recognizing that the actual value of data lies not in its commercial value but in the services and transparency it can provide to the public, governments must shift their focus towards facilitating widespread access. While embracing this approach, Latin American governments will demonstrate a commitment to democratic principles, ensuring that the wealth of information generated by private and public institutions is a shared resource, enhancing collaboration and, ultimately, leading to more effective, responsive, and inclusive governance.

\textbf{Multi-stakeholders’ commitment.} Non-government Latin American actors still require to be integrated since the development of OGDIs has traditionally been conducted primarily by government institutions. A shared vision of open data could be instrumental to operationalizing stakeholders and citizen collaboration, although this might be challenged by institutional governance \cite{aguerre_open_2024}. Stronger links could build a critical capacity for geographic applications within open data communities \cite{sieber_open_2019}. Multi-stakeholders, especially non-government actors, need to be motivated to participate in and contribute to developing and implementing the OGDI. They could be involved in both the governance and implementation, as well as the availability of various instruments could encourage this participation, including a shared vision and a plan for opening geographical data and OGDI \cite{vancauwenberghe_assessing_2018}. To prompt government action, stakeholders should advocate for policies that promote OGDIs, clearly demonstrating their societal benefits through data-driven case studies. Moreover, a comprehensive open data directive encompassing all geographical data and a management system that includes representation from non-government entities could also play a role \cite{vancauwenberghe_assessing_2018}. 

\textbf{Transitioning GDIs to OGDIs with a multi-stakeholder focus}. OGDI can be boosted by transitioning GDIs to OGDIs, which raises the opportunity, on the one hand, to build a community and foster collaboration among GDI experts, government and non-government sectors to share experiences, knowledge and expertise for policy formulation and, on the other hand, to learn from international experiences \cite{aguerre_open_2024}. For a successful transition, it is essential to clearly assign roles and responsibilities and define which institutions will be in charge of developing and implementing such policies \cite{barenque_implementacion_2021}. Also, the teams in charge of guidance and implementing the open data policies require a consistent and sustainable investment. Furthermore, besides formal, legal, and organizational structures, trained personnel and people involved in the implementation and awareness of the concept and its benefits are also needed \cite{BRUZZA_2020, martinez_analysis_2023, restrepo_scoping_2023}. An impulse for institutional enablers is needed to overcome the high heterogeneity among countries in this regard (Figure 1, GovTech Enablers Index - GTEI) \cite{world_bank_govtech_2022}. 

\subsection{Impact dimension}
\textbf{Monitoring impact}. Since rapid OGDIs development is being pursued for the region, there is a need to design and implement evaluation strategies to observe the impacts and benefits at large and in a continuous mode. These strategies need to be built locally within Latin America as well as integrated into the local institution models \cite{jacquin_citizen_2023}, opening the door to develop a geographical information science observatory \cite{claramunt_geographical_2020}. From this observatory, qualitative and quantitative metrics and models should be developed to assess a given OGDI, the collected open data, and what kind of functionalities and services, at each level, should be evaluated and adapted to the region. This observatory will also be required to identify the main geographical data involved and their management practices, as well as the kinds of communities involved (experts, decision-makers, citizens) and for what sorts of territorial practices and applications \cite{jacquin_citizen_2023}.

\textbf{Impact within the data value chain}. Not only an OGDI should qualify the data used but also the processes under which the data is used (methods and processes to secure the most significant impact at the citizen level). The impact dimension needs to be added within the data value chain itself. For instance, by adding to the collection and publishing stages - both being the two usual, well-known stages - an uptake stage (connect data and users; incentivize users to make informed decision-making; and promote a data use culture) and an impact stage (relate data and problem-decision processes; combine data sources and re-use them, track behavior and situation changes) \cite{open_data_watch_2018}.

\textbf{Geographical data literacy.} Although expert geographical data literacy has grown, the availability of tools and resources to promote data literacy in society is still limited. The current lack of proficiency in society at large represents a critical bottleneck to the effective use of open geographical data \cite{sieber_open_2019}. This issue is highlighted in Latin America because of digital divides across the whole region. Establishing and maintaining OGDIs hinges on the intricate task of fostering community engagement and collaboration and should largely rely on successful initiatives \cite{jacquin_citizen_2023}. This will require the development of an appropriate education curriculum at different levels of expertise. Low-cost and low-technical (easy to use) are specific directions to explore and discuss \cite{jacquin_citizen_2023}, as well as their potential, standards, and diffusion/acceptance across different regional and user contexts.

\textbf{Citizen engagement}. Implementing an open data model fosters a more informed and engaged citizenry \cite{martinez_estudio_2022-1}, promoting accountability and enabling innovation. By making public datasets openly available, governments empower researchers, businesses, and citizens to derive insights that contribute to societal progress. Following the principles of citizen science, end-users should be involved at the early stage of OGDIs, so this is an open and vital research challenge \cite{jacquin_citizen_2023}. A key question is how best to engage them in this process (participation, gamification, reward). The availability of public participation platforms and citizen feedback mechanisms is essential to guarantee citizen engagement and boost the benefits; however, so far, citizen engagement is uneven in the region – see Figure 1 Digital Citizen Engagement Index - DCEI \cite{world_bank_govtech_2022}. As the main geographical data generators, governments should redouble their efforts to include the business sector and citizens in the data open ecosystem to advance the agenda and generate better and broader data uses to benefit various segments of society \cite{martinez_analysis_2023}.

\section{Conclusions}

This study showcased the Latin American region by analyzing the recent progress in developing OGDIs and identifying the main challenges. This study did not seek to provide an exhaustive review of Latin American open data applications but rather to unveil the region's current OGDI landscape. Therefore, we provide practical and methodological challenges for fostering innovation, citizen engagement, digital transformation, and data-driven initiatives under the open geographical data umbrella.

While open data initiatives are growing in the region, the amount of open geographical datasets and initiatives is relatively modest. The early (or delayed) development of OGDI in the region opens the doors for a timely involvement of citizens and non-government sectors to share needs, experiences, knowledge, and expertise and adds this topic to the Latin America research agenda for geographical data. The ongoing development of the open data community allows for anticipating challenges and learning from this community experience. This learning process is essential for understanding both the successes and shortcomings of current implementations. While still in the early stages of development, this favors a readiness for formulating and adopting open geographical data policies and innovation. 

There is still a need for a minimum level of standardization of common practices for open geographical data that consider the diversity of regional, socio-cultural, and application differences. Indeed, the rapid proliferation of OGDI might lead to inconsistencies and challenges in establishing standardized practices, especially when projects lack adequate planning and consideration of existing frameworks. Therefore, every effort should be made to derive a shared common knowledge and favor capacity building. Governance structures and funding, maintenance mechanisms and protocols, and legal, ethical, and security aspects should also be defined and maintained in the long term to support ongoing data management and infrastructure development. Validation processes and quality measures ensure that the shared data adhere to predefined standards and are reliable for decision-making processes. This is a critical objective to enhance successful citizens’ participation and the reliability of OGDIs in the long term, as evidenced by various global initiatives that highlight the importance of such frameworks. This paper intentionally excluded a discussion of the FAIR principles to maintain a focused scope, although the connection between OGDI, FAIR, and open research is acknowledged. Surely this issue warrants further exploration in future studies.

It is important to note that addressing these challenges should not occur in isolation from one another despite having been identified in different dimensions. Instead, they should be interconnected and implemented cross-dimensionally. For example, establishing clear protocols, methodologies, and policies will enable the integration of citizens, business, and non-governmental actors for data generation and utilization. Similarly, enhancing geographical data literacy will promote open data availability and facilitate the adoption of innovative technologies and the assessment of their impact.

OGDIs could play a significant role in transparency in government operations and decision-making. Thus, ethics, data governance, bureaucratic barriers and security issues should be addressed collectively to foster a more inclusive and effective open data ecosystem. The digital divide among and within countries imposes an extra challenge for widely achieving a homogeneous advance in OGDI and citizen participation from different sectors. 

Communication and interaction between different regional OGDIs should be a significant direction for exchanging experiences and practices as well as data for boundary-bridging between different disciplines. There is an opportunity for the region to develop an observatory to evaluate options for implementing theoretical, methodological, and technologically innovative frameworks \cite{claramunt_geographical_2020}. By instrumenting such observatories, the aim is to provide better opportunities for cross-fertilization within the field and to improve the potential of geographical information science in research, academia and at large for the general public. The rationale for this proposal relies on the already well-established GDIs in the region accompanying (non-thematic specificity) open policies as well as the already built experience on citizen and non-government integration made by open data communities. Regional and transnational organizations can play a role in developing and supporting good practices and capacity-building for transitioning from GDIs to open ecosystems. 

\section*{Acknowledgments}

We would like to acknowledge the organizers of the 4th GIS Latin America conference held in Mexico City in September 2023 for fostering a platform to discuss Geographic Information Science and education. Their initiative sparked most of the reflections we then pursued on advancing OGDIs in Latin America and addressing both progresses made and challenges ahead, as evidenced in this paper. The authors extend their sincere thanks to the reviewers for their insightful and constructive comments, which have been instrumental in improving the quality of the manuscript. 

\begingroup
\raggedright 
\setlength{\parindent}{0pt} 
\bibliographystyle{josisacm}
\bibliography{OGDI}
\endgroup

\end{document}